# ESTIMATION OF SEVERITY OF SPEECH DISABILITY THROUGH SPEECH ENVELOPE

Anandthirtha B. GUDI[1], H. K. Shreedhar[2] and H. C. Nagaraj[3]

[1, 2] Department of Electronics and Communication Engineering,
Sri Bhagawan Mahaveer Jain College of Engineering (SBMJCE), Bangalore-562112,
Karnataka, India
[1] gudi_anand@rediffmail.com , [2] shreedhar_hk@yahoo.com
[3] Department of Electronics and Communication Engineering,
Nitte Meenakshi Institute of Technology (NMIT), Yelahanka, Bangalore-560064,
Karnataka, India
[3] principal@nmit.ac.in

## ABSTRACT

*In this paper, envelope detection of speech is discussed to distinguish the pathological cases of speech disabled children. The speech signal samples of children of age between five to eight years are considered for the present study. These speech signals are digitized and are used to determine the speech envelope. The envelope is subjected to ratio mean analysis to estimate the disability. This analysis is conducted on ten speech signal samples which are related to both place of articulation and manner of articulation. Overall speech disability of a pathological subject is estimated based on the results of above analysis.*

## 1. INTRODUCTION

Speech is one of the very important medium of communication. The quality of speech must be such that the information content must be easily be understood by human listeners. Speech disability is one of the widely encountered disabilities in children. In this direction, many researchers have tried different methods to analyze the speech using different envelope techniques like autocorrelation envelope, temporal envelope, power envelope and spectral envelope. A new speech measure based on parameterization of the autocorrelation envelope of the AM response is developed for vocal fold pathology assessment [11]. Voice conversion system is created to change the perceived speaker identity of a speech signal, which is based on converting the LPC spectrum and predicting the residual as a function of the target envelope parameters [1]. An algorithm is developed to detect speech pauses by adaptively tracking minima in a noisy signal's power envelope both for the broadband signal and for the high-pass and low-pass filtered signal [17]. A method based on the temporal envelope representation of speech is developed to reflect the perceptual characteristics of human auditory system and human speech production system [8]. The estimate of the spectral envelope by linear prediction has sharp peaks for high-pitch speakers. To overcome this, regularized linear prediction is applied to find a better estimate of the spectral envelope [14].All pole spectral envelope estimation for speech signals is addressed by establishing a band limited interpolation of the observed spectrum by using a recently rediscovered true envelope estimator and then using the band limited envelope to derive an all pole envelope model. [9]. It is confirmed that a speaker's vocal individuality is contained in the inter-band correlations of narrow-band temporal envelopes. Two types of envelope correlation matrices (ECMs) were made, using three utterances of an identical sentence. Type-A (reference) ECMs of two of the utterances were constructed to make a speaker's individual template and a type-B ECM was constructed using the other utterance. Speaker matching tests between the two types of ECMs were conducted to verify the validity of the individual speakers [19].In this paper, an attempt is made to distinguish the cases of pathological subjects with respect to normal subject by using speech envelope technique. In this present work, the speech uttered by children of age group of 5 to 8 years was recorded and digitized. The digitized signal was further processed by using a high level computational platform MATLAB.

## 2. METHODOLOGY

The study was conducted on children of native speakers of Kannada language. Prior to the data collection, teachers were requested to train the normal children aged between 5 and 8 years, to read a few words/sentences. The sentences read by children were recorded using Intelligent Portable Occular Device (IPOD) in digital form. The recording was carried out in a pleasant atmosphere. The recorded signal is transformed into .wav file by using GOLDWAVE software. The same procedure is also followed while recording the speech samples of children with speech disorder. The data was collected at J.S.S. 'Sahana' integrated and special school located at Bangalore, India.

Energy level for a particular recorded word "Namma" pronounced by 60 normal children was found and average energy level is calculated. The energy level of the child which is very close to average energy level is considered as Normal Subject. The child with speech disability is considered as a Pathological Subject.

The speech disorder [1, 8] depends on the problems associated with either place of articulation [15] or manner of articulation or both. The specific letters shown in the Table-1 may not be enough to study the speech disorder completely. Words with more than two letters result in greater computational complexity in computing speech disorder. Hence, in the present study, emphasis is given for words with two letters particularly ending with a specific letter shown in the Table-1.

Table-1. List of specific Kannada letters/ words

| Sl No. | Place of articulation | Manner of articulation | Kannada Letters | words in Kannada language |
|---|---|---|---|---|
| 1 | Bilabial | Nasals | ಮ(m) | ನಮ್ಮ (Namma) |
| 2 | Dental | Nasals | ನ (n) | ನನ್ನ (Nanna) |
| 3 | Dental | Fricatives | ತ(t) , ದ(d) | ಇದೆ (Ide) |
| 4 | Alveolar | Laterals | ಲ(l) , ಳ (L) | ಶಾಲೆ (Shale) |
| 5 | Bilabial | Stops | ಪ (p) , ಬ(b) | ಜೇಪಿ (Jepi) |

| 6 | Palatal | Affricates | ಜ(j) , ಸ(s) ಶ(sha) | ಹೆಸ(Hesa) |
|---|---|---|---|---|
| 7 | Velar | Plosives | ಕ(k), ಗ(g) | ನಗ(Naga) |
| 8 | Glottal | | ಹ(h) | ಸಹ( Saha) |
| 9 | | Trills | ರ(r) | ದೂರ(Doora) |
| 10 | | Glides | ಯ(y) | ಜಯ(Jaya) |

## 3. DETERMINATION OF ENVELOPE OF SPEECH SIGNAL

The Recorded speech signals samples for common Kannada word "Namma" uttered by normal subject is considered. Start and end points of this word is determined. The digitized speech signal consists of total 21121 samples. The sample count is rounded down to nearest hundred by removing the trailing samples. The speech signal is divided into multiple blocks with block size of 100 samples [5]. In each block, maximum and minimum values of the samples are determined. The speech envelope of this block is determined by linearly interpolation [5] of samples in between maximum and minimum values. The same procedure is applied for all the blocks to determine the complete envelope of the uttered word "Namma". The speech signal and the corresponding envelope is indicated in Figure 1.

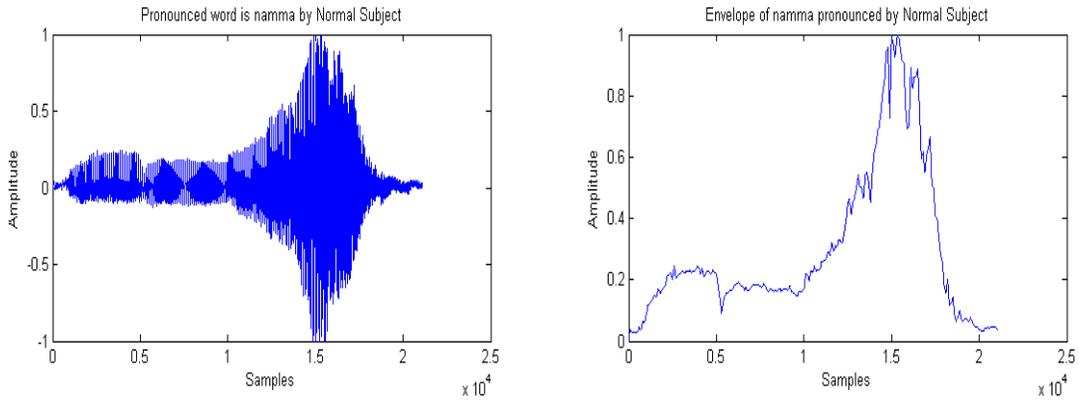

Figure 1. Amplitude plot and Envelope of the word "Namma" of Normal Subject

## 4. DETERMINATION OF THRESHOLD

To calculate the threshold value, three utterances of the word "Namma" by Normal Subject were considered. Envelope of each of these samples is calculated. Envelope of first utterance is treated as the reference. Ratio of absolute value of envelope of second utterance and first utterance and that of third and first utterance is calculated. Mean values of the ratios are calculated. Out of these two values, minimum value is chosen as threshold value.

## 5. ESTIMATION OF SPEECH DISABILITY

Speech signal samples of the uttered word "Namma" by three pathological subjects are considered. The length of samples is equated to that of normal subject. Envelopes of these utterances are determined. For each pathological subject, mean of absolute ratio between envelope of pathological subject and envelope of normal subject is calculated. This analysis method is repeated for all the words indicated in Table-1. The speech signal samples of the word "Namma" uttered by each of three pathological subjects and their corresponding envelopes are shown in Figure 2, Figure 3 and Figure 4.

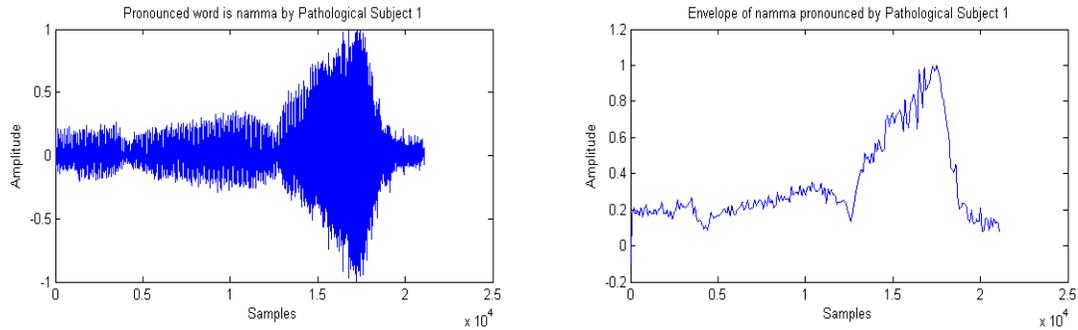

Figure 2. Amplitude plot and Envelope of the word "Namma" of Pathological Subject 1

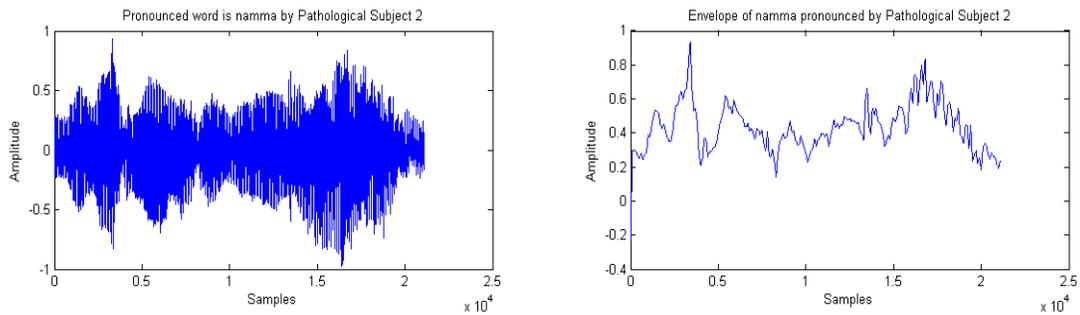

Figure 3. Amplitude plot and Envelope of the word "Namma" of Pathological Subject 2

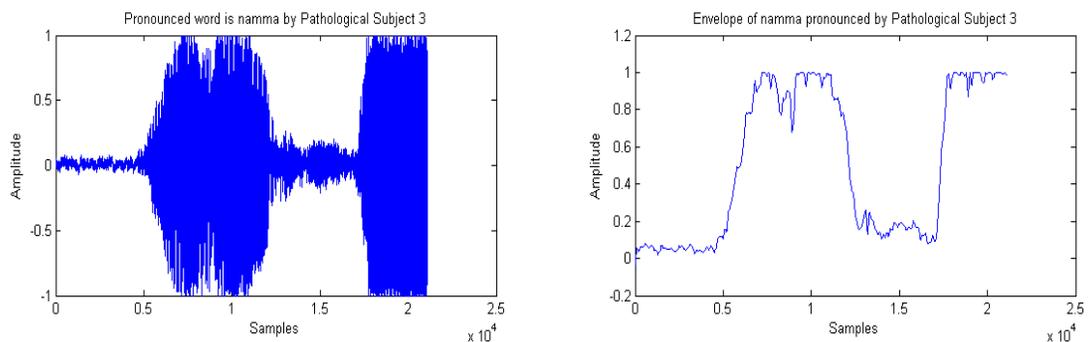

Figure 4. Amplitude plot and Envelope of the word "Namma" of Pathological Subject 3

The limits used for the classification of mild, moderate and severe cases of speech disorders of Pathological Subjects are indicated in Table-2. For this classification, threshold value of 1.03989 is used. This value is obtained from envelope analysis of Normal Subject.

Table-2 Limits for assessment of varying degree of speech disability

| No Deviation | Small Deviation | Moderate Deviation | Large Deviation |
|---|---|---|---|
| $0.8<R_{th}<=1.2$ | $1.2<R_{th}<=1.8$ or $0.6<=R_{th}<0.8$ | $1.8<R_{th}<=2.6$ or $0.4<=R_{th}<0.6$ | $R_{th}>2.6$ or $R_{th}<0.4$ |

## 5. EXPERIMENTAL RESULTS

Table-3 indicates the result of analysis of three pathological subjects for ten words. The entire analysis is conducted by using MATLAB tool. The summary of the above analysis is represented in Table-4.

Table-3. Result of analysis of various words for the three Pathological Subjects

| Pronounced Word | Particulars | Ratio Value | Deviation from the Threshold Value |
|---|---|---|---|
| Namma | Normal and Pathological Subject 1 | 1.62684 | mild |
|  | Normal and Pathological Subject 2 | 2.60459 | moderate |
|  | Normal and Pathological Subject 3 | 4.22875 | severe |
| Ide | Normal and Pathological Subject 1 | 0.94963 | no |
|  | Normal and Pathological Subject 2 | 0.45808 | moderate |
|  | Normal and Pathological Subject 3 | 1.35224 | mild |
| Shale | Normal and Pathological Subject 1 | 1.54192 | mild |
|  | Normal and Pathological Subject 2 | 1.70836 | mild |
|  | Normal and Pathological Subject 3 | 2.19511 | moderate |
| Jepi | Normal and Pathological Subject 1 | 2.30208 | moderate |
|  | Normal and Pathological Subject 2 | 3.52232 | large |
|  | Normal and Pathological Subject 3 | 3.19259 | severe |
| Hesa | Normal and Pathological Subject 1 | 1.39495 | mild |
|  | Normal and Pathological Subject 2 | 1.66019 | mild |
|  | Normal and Pathological Subject 3 | 3.23197 | severe |
| Naga | Normal and Pathological Subject 1 | 1.04994 | no |
|  | Normal and Pathological Subject 2 | 2.05953 | moderate |
|  | Normal and Pathological Subject 3 | 1.09738 | no |
| Saha | Normal and Pathological Subject 1 | 1.67477 | mild |
|  | Normal and Pathological Subject 2 | 1.43331 | mild |
|  | Normal and Pathological Subject 3 | 2.22832 | moderate |
| Doora | Normal and Pathological Subject 1 | 0.76715 | mild |
|  | Normal and Pathological Subject 2 | 1.47577 | mild |
|  | Normal and Pathological Subject 3 | 1.42325 | mild |
| Jaya | Normal and Pathological Subject 1 | 1.20300 | no |
| | Normal and Pathological Subject 2 | 3.35342 | large |
| | Normal and Pathological Subject 3 | 2.77495 | severe |
| Nanna | Normal and Pathological Subject 1 | 1.51143 | mild |
| | Normal and Pathological Subject 2 | 2.34976 | moderate |
| | Normal and Pathological Subject 3 | 2.96658 | severe |

Table-4. Summary of Results

| Name of the Subject | No Deviation | Mild Deviation | Moderate Deviation | Large Deviation |
|---|---|---|---|---|
| Pathological Subject 1 | 3 | 6 | 1 | Nil |
| Pathological Subject 2 | Nil | 4 | 4 | 2 |
| Pathological Subject 3 | 1 | 2 | 2 | 5 |

From the Table-4, it can be concluded that the overall severity of speech disability is mild for Pathological Subject 1, moderate with Subject 2 and severe with Subject 3.

## 6. CONCLUSIONS

The present study presents a method to estimate the severity of speech disability of a speech disabled child by using the speech signal samples. With this, it is possible to detect speech disorder in children at early stages. Hence necessary corrective measures can be taken. The analysis made in the present study helps the speech therapist for finding the degree of improvement in the speech disabled child by analyzing periodically recorded speech samples.

## ACKNOWLEDGEMENTS

The authors would like to thank the management of Sri Bhagawan Mahaveer Jain College of Engineering, Bangalore, Nitte Meenakshi Institute of Technology, Bangalore, and J.S.S. 'Sahana' Integrated and special school, Bangalore for their Constant support and encouragement in undertaking the research work.

## REFERENCES

[1] Alexander Kain and Michael W Macon, "Design and Evaluation of a Voice Conversion Algorithm Based on Spectral Envelope Mapping and Residual Prediction", 2001 IEEE, pp 813-816.

[2] Alireza A. Dibazar, S. Narayanad and T. W. Berger, "Feature Analysis for Automatic Detection of Pathological Speech", Proceedings of the Second Joint EMBSBMES Conferenw Houston, TX, USA October 23-26, 2002, pp 182-183.

[3] Anandthirtha. B. Gudi and H. C. Nagaraj, "Speech Disability Threshold Determination by Graphical and DSP Techniques", Sensors & Transducers Journal, Vol. 107, Issue 8, August 2009, pp. 157-164

[4] Anandthirtha B Gudi, Shreedhar H.K and Dr. H. C. Nagaraj, "Difference Ratio Analysis of Speech to Estimate the Severity of Speech Disability", 2011 3rd International Conference on Computer Modeling and Simulation (ICCMS 2011).

[5] Ben Gold and Nelson Morgan, "Speech and Audio Signal Processing (Processing and Perception of Speech and Music)", John Wiley & Sons Inc., 2006.

[6] Bishwarup Mondal and T V Sreenivas, "Mixture Gaussian Envelope Chirp Model for Speech and Audio", IEEE, 2001, pp 857-860.

[7] CHERIF Adnbne-Botiafif Lamia-Mhamdi Mounir, "Analysis of Pathological Voices by Speech Processing", Sensors & Transducers Journal, Vol. 107, Issue 8, August 2009, pp. 157- 164 IEEE, 2003, pp. 365-367.

[8] Doh-Suk Kim, "A Cue for Objective Speech Quality Estimation in Temporal Envelope Representations" IEEE signal processing letters, Vol. 11, No. 10, October 2004, pp 849 -852.

[9] Fernando Villavicencio, Axel Röbel and Xavier Rodet, "Improving LPC Spectral Envelope Extraction of Voiced Speech by True-Envelope Estimation", IEEE, ICASSP 2006, pp 869-872.

[10] Guojun Zhou, John H. L. Hansen and James F. Kaiser, "Nonlinear Feature Based Classification of Speech under Stress", IEEE Transactions on Speech and Audio Processing, vol. 9, No. 3, March 2001, pp 201-216.

[11] John H. L. Hansen, Liliana Gavidia-Ceballos, and James F. Kaiser, "A Nonlinear Operator-Based Speech Feature Analysis Method with Application to Vocal Fold Pathology Assessment", IEEE Transactions on Biomedical Engineering, Vol. 45, no. 3, March 1998, pp 300-313.

[12] Karthikeyan Umapathy, Sridhar Krishnan, Vijay Parsa and Donald G. Jamieson, "Discrimination of Pathological Voices Using a Time-Frequency Approach", IEEE Tansactions on Biomedical Engineering, Vol. 52, No. 3, March 2005, pp 421-430.

[13] Kyung-Tae Kim, Min-Ki Lee and Hong-Goo Kang, "Speech Bandwidth Extension Using Temporal Envelope Modeling", IEEE Signal Processing Letters, Vol. 15, 2008, pp 429-432.

[14] L. Anders Ekman, W. Bastiaan Kleijn and Manohar N. Murthi, " Spectral Envelope Estimation and Regularization" – IEEE, ICASSP -2006 pp.245-248.

[15] Lawrence Rabiner and Biling-Hwang Juang, Fundamentals of Speech Recognition, second Impression, Dorling Kindersley (India) Pvt. Ltd., licensees of Pearson Education in South Asia. 2007.

[16] Maria Markaki and Yannis Stylianou,"Using Modulation Spectra for Voice Pathology Detection and Classification", 31st Annual International Conference of the IEEE EMBS Minneapolis, Minnesota, USA, September 2-6, 2009, pp 2514-2517.

[17] Mark Marzinzik and Birger Kollmeier, "Speech Pause Detection for Noise Spectrum Estimation by Tracking Power Envelope Dynamics", IEEE transactions on speech and audio processing, Vol. 10, No. 2, February 2002, pp. 109-118.

[18] Martinez Cdsar E and Rufiner Hugo L, "Acoustic Analysis of Speech for Detection of Laryngeal Pathologies" Proceedings of the 22nd Annual EMBS International Conference, July 23-28, 2000, Chicago IL, pp2369-2372.

[19] S. Gotoh, M. Kazama, M. Tohyama and Y. Yamasaki, "Speaker Verification Using Narrow-band Envelope Correlation Matrices" 2006 IEEE International Symposium on Signal Processing and Information Technology pp-310-313.

[20] Sriram Ganapathy Samuel, Thomas and Hynek Hermansky, "Temporal Envelope Subtraction for Robust Speech Recognition Using Modulation Spectrum", IEEE, ASRU 2009, pp 164-169.

Signal & Image Processing : An International Journal (SIPIJ) Vol.2, No.2, June 2011

[21] Steven Schimmel and Les Atlas, "Coherent Envelope Detection for Modulation Filtering of Speech",IEEE, ICASSP 2005, pp 221-224.

[22] Wieslaw Wszd-ek, Ryszard Tadeusiewicz, Andrzej Izworski and Tadeusz Wszolek, "Automated Understanding of Selected Voice Tract Pathologies Based on the Speech Signal Analysis", 2001 Proceedings of the 23rd Annual EMBS International Conference, October 25-28, Istanbul, Turkey, pp 1719-1722.

**Biographies:**

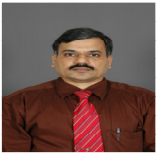

**Anandthirtha B. Gudi** obtained Bachelor of Engineering from S.J.M.Institute of Technology, Chtradurga, Mysore University**,** Master of Engineering from U.V.C.E, Bangalore University**.** Bangalore Professor in the Department of Electronics and Communication Engineering, Sri Bhagawan Mahaveer Jain College of Engineering, Bangalore, Karnataka, India.

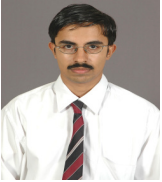

**H.K.Shreedhar** obtained Bachelor of Engineering from U.V.C.E. Bangalore, Bangalore University. Master of Technology from B.M.S.C.E., Bangalore, V.T.U. Assistant Professor in the Department of Electronics and Communication Engineering, Sri Bhagawan Mahaveer Jain College of Engineering, Bangalore, Karnataka, India.

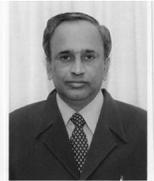

**Dr. H. C. Nagaraj** obtained Bachelor of Engineering from Mysore University**,** Master of Engineering from P.S.G.College of Technology, Coimbatore. Ph.D from IIT.Madras. Principal, Nitte Meenakshi Institute of Technology, Bangalore, Karnataka, India.